\documentclass[aps,twocolumn,showpacs]{revtex4-1}
\usepackage{graphicx}
\usepackage[all]{xy}
\usepackage{amsmath}
\usepackage{amssymb}
\newcommand{\be}{\begin{equation}}
\newcommand{\ee}{\end{equation}}
\newcommand{\ben}{\begin{eqnarray}}
\newcommand{\een}{\end{eqnarray}}
\newcommand{\bes}{\begin{subequations}}
\newcommand{\ees}{\end{subequations}}

\newcommand{\bb}{\bibitem}
\begin{document}
\title{New scalar field models and their defect solutions}
\author{D. Bazeia,$^a$ M.A. Gonz\'alez Le\'on,$^b$ L. Losano,$^a$ and J. Mateos Guilarte$^c$}
\affiliation{{\small $^a$Departamento de F\'{\i}sica, Universidade Federal da Para\'{\i}ba, 58051-970 Jo\~ao Pessoa, Para\'{\i}ba, Brazil\\
$^b$Departamento de Matem\'atica Aplicada and IUFFyM, Universidad de Salamanca, Spain\\
$^c$Departamento de F\'{\i}sica Fundamental and IUFFyM, Universidad de Salamanca, Spain}}

\begin{abstract}
In this work we introduce new scalar field models and study the defect solutions they may engender.
The investigation is based on the deformation procedure, which greatly simplify the calculations, leading us to
new models together with the corresponding solutions in a simple and powerful manner.
\end{abstract}


\maketitle

{\bf Generalities. --} Topological structures \cite{MS} are of general interest in physics because they may be used in a diversity of situations, ranging from high energy physics, where they can appear during phase transitions governed by spontaneous symmetry breaking \cite{VS}, going down to low energy physics, where they can contribute to pattern formation \cite{WA} and to the construction of logical apparatus at the nanometric scale \cite{DW}.

In this work we focus on new models described by real scalar fields and their topological solutions. The investigation is of current interest to a diversity of scenarios both in high energy physics and in condensed matter. In high energy physics, they may help understanding interesting problems related to topological defects; see, e.g., Refs.~\cite{dunne,sal}. In condensed matter, we can use the models to investigate interesting issues, for instance, to describe polarons in carbon nanotubes \cite{rio} and to study nonlinear excitations in Bose-Einstein condensates \cite{bec,abc}. By addressing a much broader class of potentials we hope to provide analytical basis to the more and more exotic structures arising in the areas of physics previously mentioned. 

Here we will take advantage of the deformation procedure introduced in Ref.~{\cite{blm}} for real scalar fields in the standard case, dealing with systems defined in the real line. We consider models described by the Lagrange densities
\bes
\ben
{\cal L}=\frac12\partial_\mu\chi\partial^\mu\chi-U(\chi)\label{1a},
\\
{\cal L}_d=\frac12\partial_\mu\phi\partial^\mu\phi-V(\phi)\label{1b},
\een\ees
where $\chi$ and $\phi$ are two real scalar fields and $U(\chi)$ and $V(\phi)$ are given potentials, which specify each one of the two models. The approach requires that we introduce a function $\chi=f(\phi),$ named deformation function, from which we link the model (\ref{1a}) with the deformed model (\ref{1b}). This is done relating the two potentials $U(\chi)$ and $V(\phi)$ in the very specific form
\be
V(\phi)=\frac{U(\chi\to f(\phi))}{(df/d\phi)^2}.
\ee
The choice leads to the result that if the original model has finite energy static solution $\chi(x)$ which obeys the equation of motion
\be
\frac{d^2\chi}{dx^2}=\frac{dU}{d\chi},
\ee
then the deformed model has finite energy static solution given by $\phi(x)=f^{-1}(\chi(x)),$ which obeys
\be
\frac{d^2\phi}{dx^2}=\frac{dV}{d\phi}.
\ee
The proof of the above result can be found in Ref.~{\cite{blm}}, so we omit it here.

We notice that if the deformation function $f(\phi)$ has critical points, the deformation procedure works smoothly if and only if the potential $U(f(\phi))$ has  zeros of multiplicity at least two at those critical points. In this case, the critical points of $f(\phi)$ are the branching points of $f^{-1}(\chi),$ which is consequently a multivalued function; see, e.g., Refs.~\cite{bglm,d1,d2} for other issues concerning the deformation procedure.

{\bf New models. --} We now proceed to one important point of the present work, which is to find new models and their defect solutions. To do this, we start with polynomial and irrational potentials, and we search for the corresponding kinks and/or lumps, which are the classical static solutions of the corresponding equations of motion. We follow the standard methodology, taking as the starting model a family of models, described by the potential
\be\label{upot}
U(\chi)=\frac12\;(1-\chi^2)^n.
\ee
Here we use dimensionless field and coordinates, and we take $n$ as integer or half integer, such that $n\geq 3/2$.

This is the new family of models which we consider, and for $n=2$ we get to the standard fourth order power potential with spontaneous symmetry breaking. If we choose the center of the defect at the origin, we get to the kink $(+)$ and antikink $(-)$ solutions $\pm\chi_n(x)$, where $\chi_n(x)$ can be expressed in terms of the inverse of the Gauss Hypergeometric function, in a way such that
\be\label{Fn}
_2{F}_1\left(\frac12,\frac{n}2,\frac32,\chi^2\right)\chi=x.
\ee
In the case of $n=2$ and $n=3$, the above expression is easily invertible to give the explicit solutions
\be
\chi_2(x)=\tanh(x)\;,
\ee
and
\be
\chi_3(x)=\frac{x}{\sqrt{1+x^2}}\;,
\ee
respectively. The case $n=2$ is well known and $n=3$ is nonstandard, but it was already considered in Ref.~\cite{blm}.

In general, it is very hard to write the explicit solution analytically, but the implicit plot can be given; see, e.g., Fig.~1 for the cases of $n=3/2,5/2,$ and $3$.
To go further on this, we can write the potential in the form
\be
U(\chi)=\frac12\left(\frac{d{W_n}}{d\chi}\right)^2
\ee
where
\be\label{wn}
{W_n}(\chi)= \;_2{F}_1\left(\frac12,-\frac{n}2,\frac32,\chi^2\right)\chi.
\ee
Note that formula (9) is well defined on the reals for every $\chi\in(-\infty,\infty)$ only if $n$
is an even integer. If $n$ is integer but odd $U$ is always real but $W$ is purely imaginary for $|\chi|>1$.
Finally, if $n$ is half-integer $U$ and $W$ in formula (9) are real only if $\chi\in[-1,1]$.
In any case, this leads to the energy of the static solution
\be\label{eny}
E_n=\left|{W_n}(\chi=1)-{W_n}(\chi=-1)\right|=\left|B\left(\frac12,1+\frac{n}2\right)\right|.
\ee
where $B(a,b)$ stands for the Euler Beta function. For $n$ even: $n=2m$, with $m\in {\mathbb
N}$, the corresponding superpotential \eqref{wn}, reduces to a polynomial in $\chi^2$
\be
W_{2m}(\chi)\, =\, \chi \, \sum_{k=0}^{m} (-1)^k \,
{m\choose k} \, \frac{\chi^{2k}}{2k+1}.
\ee
For $n$ odd: $n=2m+1$, the expression for the superpotential \eqref{wn} can be expressed in the form
\be
W_{2m+1}(\chi)\, =\, \frac{(2m+1)!!}{(2m+2)!!}\, \arcsin \chi +  \chi \, \sqrt{1-\chi^2} \, P_{m}(\chi^2)
\ee
where $P_{m}(\chi^2)$ is a polynomial of degree $m$ in $\chi^2$, thus of degree $2m$ in the field $\chi$.

\begin{figure}[ht]
\vspace{1cm}
\includegraphics[{height=03cm,width=6cm,angle=00}]{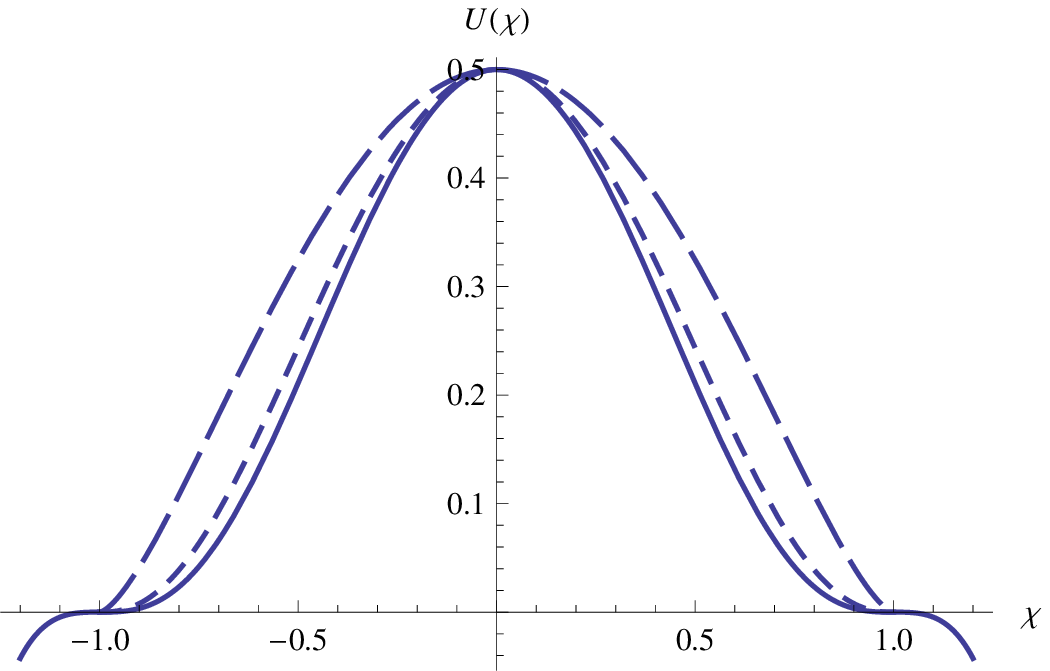}
\includegraphics[{height=03cm,width=6cm,angle=00}]{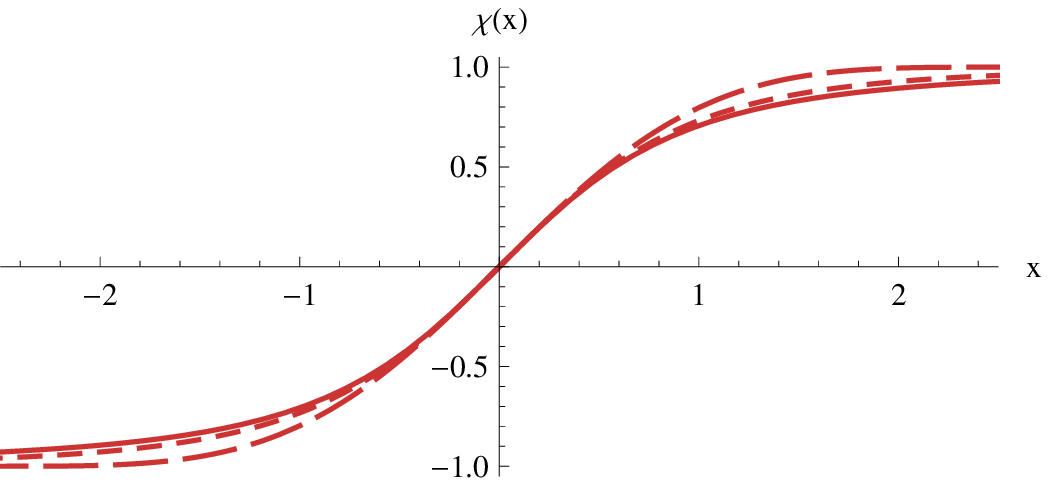}
\caption{In the upper panel: plots of the potential for $n=3/2$ (large dashing), $5/2$ (small dashing), and $3$ (continuous). In the lower panel: plots of the respective kinklike solutions.}
\end{figure}

{\bf Deformed models. --} In the previous section, we have constructed new family of models. Thus, let us now focus on the main point of this work, which concerns the construction of new models, with the use of the deformation procedure. We use the deformation function \cite{bglm}
\be\label{def}
f(\phi)=\cos(a\arccos{\phi}-m\pi),
\ee
where $a$ is a real constant, and $m$ can be integer or half integer. We deform the potential \eqref{upot}, which depends on $n$, with the deformation function \eqref{def}, which depends on $a$ and $m$. The parameter $m$ leads to two distinct families of models: for $m$ integer, the deformed potential can be written in the form
\be\label{vs1}
{V}_{\sin}^{n,a}(\phi)=\frac{1}{2a^2}{(1-\phi^2)}\sin^{2n-2}\left(a\;\arccos\phi\right)
\ee
and, for $m$ half integer we get
\be\label{vc1}
{V}_{\cos}^{n,a}(\phi)=\frac{1}{2a^2}{(1-\phi^2)}\cos^{2n-2}\left(a\;\arccos\phi\right)
\ee
The new models \eqref{vs1} and \eqref{vc1} do not depend on $m$ anymore, since we have already used the fact that $m$ is integer or half integer. Moreover,
we will see below that using integer or half integer values for the parameter $a$ leads the number of minima of the new potentials to be fixed at will.

The above models identify families of potentials which present static solutions which can be constructed with the help of the general expression
\be\label{sol1}
{\phi}(x)=\cos\left(\frac{\theta_n(x)+ m\;\pi}{a}\right)
\ee
where
\be\label{theta}
\theta_n(x)=\arccos(\chi_n(x)),
\ee
with $\chi_n(x)$ being defect solution of the model discussed in the previous section, and $\theta_n(x)\in [0,\pi]$.

Since we have a diversity of models and solutions, let us now illustrate the investigation focusing our attention on some specific cases.

{\bf New models for $a$ integer.--} Here we investigate families of models for the specific cases of $a$ and $n$ being integers. We first consider the case of $V^{n,a}_{\sin}$, as written in \eqref{vs1}. We have two possibilities:
\begin{itemize}
\item for $a$ odd:
\be
{V}_{\sin}^{n,a}(\phi)=\frac{1}{2a^2}\;(1-\phi^2)^{n} \, \prod_{j=2}^{(a+1)/2}
\left(1-\frac{\phi^{2}}{{Z_j^{a}}^2}\right)^{2n-2}
\ee
\item for $a$ even:
\be
{V}_{\sin}^{n,a}(\phi)=\frac{a^{2n-4}}{2}\;\phi^{2n-2}(1-\phi^2)^n\, \prod_{j=2}^{a/2}
\left(1-\frac{\phi^{2}}{{Z_j^{a}}^2}\right)^{2n-2}
\ee
\end{itemize}
where $Z_j^a= \cos\left(\frac{j-1}{a}\pi\right)$. The $sine$ potential can also be written in terms of the Chebyshev
polynomials of second kind, in the form
\ben
V^{n,a}_{\sin}(\phi)=\frac1{2a^2}\;(1-\phi^{2})^n\;[U_{a-1}(\phi)]^{2n-2}
\een
where
\ben
U_a(\theta)=\frac{\sin((a+1)\arccos\theta)}{\sin(\arccos\theta)}\label{cheb2}
\een
The explicit form of the potentials are given by, for $a=1,2,3$ and $n=2$
\be\label{vsin2a}
V^{2,1}_{\sin}(\phi)=\frac12\;\;(1-\phi^{2})^2
\ee
\be
V^{2,2}_{\sin}(\phi)=\frac1{2}\;\phi^2\;(1-\phi^{2})^2
\ee
\be
V^{2,3}_{\sin}(\phi)=\frac{8}{9}\;\;(1-\phi^{2})^2\left(\frac14-\phi^{2}\right)^2
\ee
and for $n=3$
\be\label{vsin3a}
V^{3,1}_{\sin}(\phi)=\frac12\;\;(1-\phi^{2})^3
\ee
\be
V^{3,2}_{\sin}(\phi)={2}\;\phi^4\;(1-\phi^{2})^3
\ee
\be
V^{3,3}_{\sin}(\phi)=\frac{128}{9}\;\;(1-\phi^{2})^3\left(\frac14-\phi^{2}\right)^4
\ee
We notice that for $n$ even, there are two classes of models: for $a$ odd they are $\phi^4-$like potentials (no zero at the origin) and for $a$ even they are $\phi^6-$like models (having a zero at the origin). And for $n$ odd, there are also two classes of models: for $a$ odd they are inverted $\phi^4-$like potentials (no zero at the origin) and for $a$ even they are inverted $\phi^6-$like models (having a zero at the origin). To illustrate the results, in Fig.~2 we plot some of the potentials as a function of the scalar field $\phi$.

\begin{figure}[ht]
\vspace{1cm}
\includegraphics[{height=03cm,width=9cm,angle=00}]{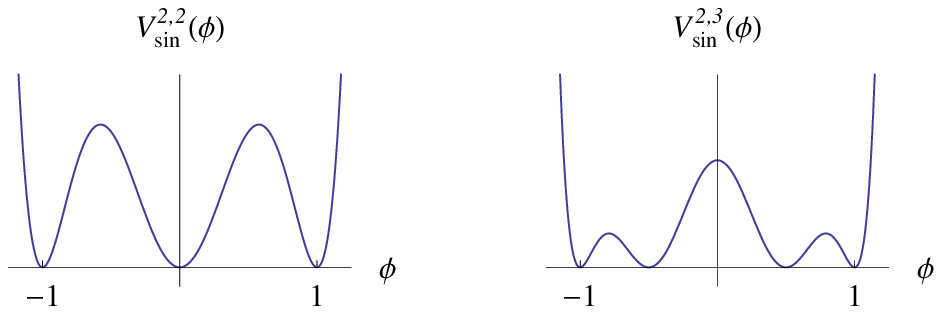}
\includegraphics[{height=03cm,width=9cm,angle=00}]{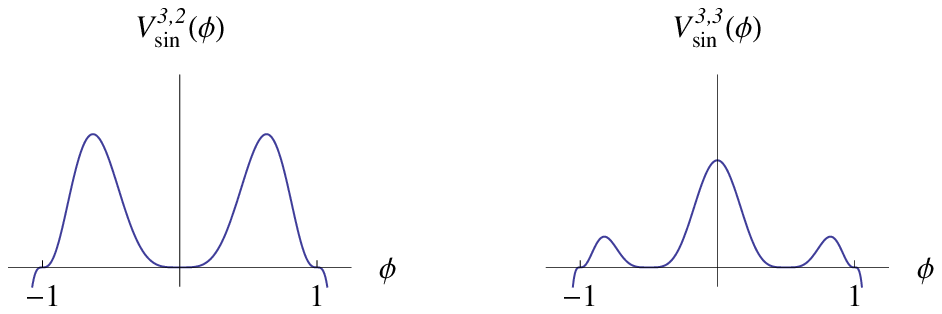}
\includegraphics[{height=03cm,width=9cm,angle=00}]{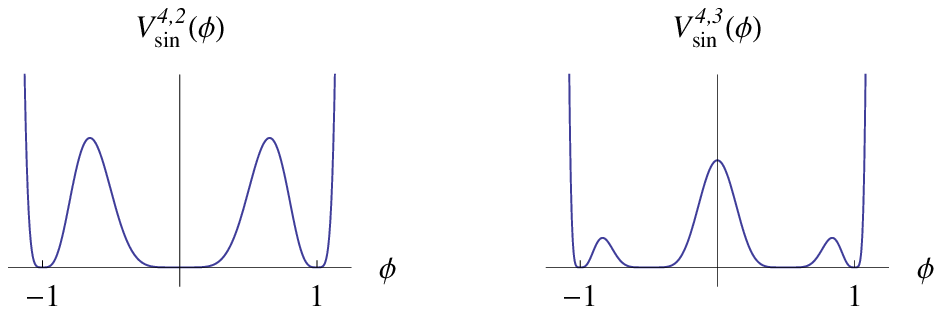}
\caption{Plots of $V^{n,a}_{\sin}$ for some $n$ and $a$.}
\end{figure}

The static solutions of the potentials $V^{n,a}_{\sin}(\phi)$ are given by
\be\label{solsin1}
{\phi}_k(x)=\cos\left(\frac{\theta_n(x)}{a}+ \frac{(k-1)\;\pi}{a}\right),
\ee
where $k$ is integer, which produces different solutions only if $1\leq k\leq 2a$; recall that $\theta_n(x)$ is given by \eqref{theta}.
We notice that, for $n$ even, the potentials of the $sine$ family are non negative, and all the zeros are minima. Then, all the static solutions given by Eq.~\eqref{solsin1} are topological, since they interpolate consecutive minima. In the case for $n$ odd, the potentials of the $sine$ family are non negative only for $|\phi|<1$, and the zeros at $\phi=\pm1$ are not minima.
Then, all the static solutions given by Eq.~\eqref{solsin1} which interpolate consecutive local minima are topological, and the others, for $k=a,2a$, which involves the zeros at $\phi=\pm1$ and the neighbor local minimum, are non topological.

Let us now consider the family of models $V^{n,a}_{\cos}$ given by \eqref{vc1}, for the specific cases of $a$ and $n$ being integers. Here we also have two possibilities:
\begin{itemize}
\item for $a$ odd:
\be
{V}_{\cos}^{n,a}(\phi)=\frac{a^{2n-4}}{2}\;\phi^{2n-2}(1-\phi^2)\, \prod_{j=1}^{(a-1)/2}
\left(\frac{\phi^{2}}{{Z_j^{a}}^2}-1\right)^{2n-2}
\ee
\item for $a$ even:
\be
{V}_{\cos}^{n,a}(\phi)=\frac{1}{2a^2}\;(1-\phi^2)\, \prod_{j=1}^{a/2}
\left(\frac{\phi^{2}}{{Z_j^{a}}^2}-1\right)^{2n-2}
\ee
\end{itemize}
where $Z_j^a= \cos\left(\frac{2j-1}{2a}\pi\right)$.
The $cosine$ potentials $V^{n,a}_{\cos}(\phi)$ can also be written in terms of Chebyshev polynomials of the first kind
\ben
V^{n,a}_{\cos}(\phi)&=&\frac1{2a^2}\;(1-\phi^{2})\;[T_a(\phi)]^{2n-2}
\een
where
\ben
T_a(\theta)&=&\cos(a\arccos\theta)\label{cheb1}
\een
The explicit form of the potentials are given by, for ${a=1,2,3}$ and $n=2$
\be\label{vcos2a}
V^{2,1}_{\cos}(\phi)=\frac12\;\phi^2\;(1-\phi^{2})
\ee
\be
V^{2,2}_{\cos}(\phi)=\frac1{2}\;(1-\phi^{2})\;\left(\frac12-\phi^{2}\right)^2
\ee
\be
V^{2,3}_{\cos}(\phi)=\frac{8}{9}\;\phi^2\;(1-\phi^{2})\left(\frac34-\phi^{2}\right)^2
\ee
and for $n=3$
\be\label{vcos3a}
V^{3,1}_{\cos}(\phi)=\frac12\;\phi^4\;(1-\phi^{2})
\ee
\be
V^{3,2}_{\cos}(\phi)=\frac1{2}\;(1-\phi^{2})\;\left(\frac12-\phi^{2}\right)^4
\ee
\be
V^{3,3}_{\cos}(\phi)=\frac{128}{9}\;\phi^4\;(1-\phi^{2})\left(\frac34-\phi^{2}\right)^4
\ee
We notice that there are two classes of models: for $a$ even they are inverted $\phi^4-$like potentials (no zero at the origin) and for $a$ odd they are inverted $\phi^6-$like models
(having a zero at the origin). To illustrate the results, in Fig.~3 we plot some potentials as a function of the scalar field $\phi$.

\begin{figure}[ht]
\vspace{1cm}
\includegraphics[{height=03cm,width=9cm,angle=00}]{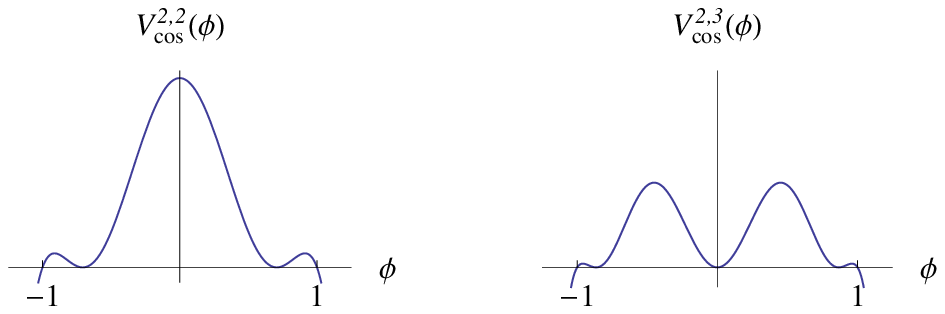}
\includegraphics[{height=03cm,width=9cm,angle=00}]{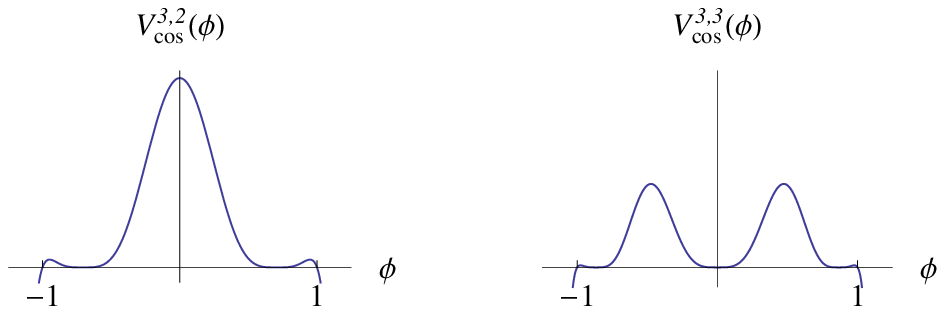}
\includegraphics[{height=03cm,width=9cm,angle=00}]{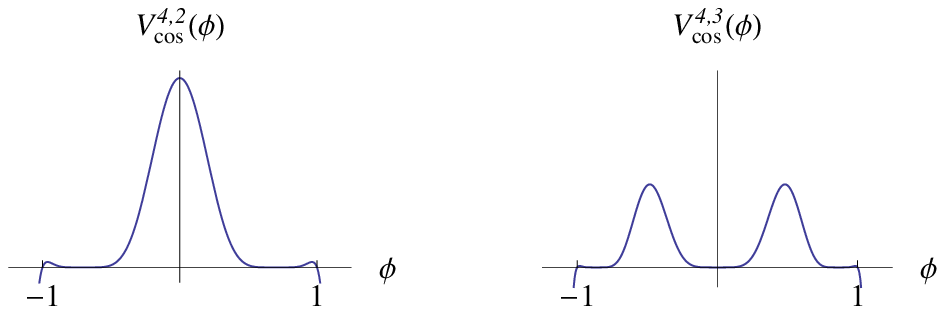}
\caption{Plots of $V^{n,a}_{\cos}$ for some $n$ and $a$.}
\end{figure}

The static solutions of the $V^{n,a}_{\cos}(\phi)$, are given by
\be\label{solcos1}
{\phi}_k(x)=\cos\left(\frac{\theta_n(x)}{a}+ \frac{(2k-1)\;\pi}{2a}\right)\;,
\ee
where $k$ is integer, which produces different solutions only if $1\leq k\leq 2a$; recall that $\theta_n(x)$ is given by \eqref{theta}.
We notice that, the potentials of the $cosine$ family are non negative only for $-1\leq\phi\leq1$, and the zeros at $\phi=\pm1$ are not minima.
Then, for $n$ integer, all the static solutions given by Eq.\eqref{solcos1} which interpolate consecutive minima are topological, and the others, for $k=a, 2a$, which involves the zeros at $\phi=\pm1$ and the neighbor minima, are non topological.

{\bf New models for $a$ half integer. --} In this case, the two families of models are field reflection of one another, that is, $V^{n,a}_{\cos}(\phi)=V^{n,a}_{\sin}(-\phi)$. Thus, they have no important difference, and we will examine explicitly only one of them. The potentials of the $sine$ family can be expressed in the form
\be\label{vsinsemi}
{V}_{\sin}^{n,a}(\phi)=\frac{1}{2^n a^2}\;(1-\phi^{2})(1-\phi)^{n-1} \;
\prod_{j=1}^{a-\frac12} \left( 1+\frac{\phi}{Z_j^a}\right)^{2n-2},
\ee
where $Z_j^a= \cos\left(\frac{2j-1}{2a}\pi\right)$, and $n$ is an integer.
They can also be given in terms of Chebyshev polynomials of the first kind
\be
V^{n,a}_{\sin}(\phi)=\frac{1}{2^n a^2}\;(1-\phi^{2})\left(1-T_{2a}(\phi)\right)^{n-1}
\ee
where $T_\alpha(\theta)$ is given by Eq.~(\ref{cheb1}).
The explicit form of the potentials are given by, for $a=1/2,3/2,5/2$ and $n=2$
\be
{V}_{\sin}^{2,1/2}(\phi)=(1-\phi^2)\;(1-\phi)
\ee
\be
{V}_{\sin}^{2,3/2}(\phi)=\frac{1}{9}(1-\phi^2)\;(1-\phi)\;(1+2\phi)^2
\ee
\be
{V}_{\sin}^{2,5/2}(\phi)=\frac{1}{25}\;(1-\phi^2)(1-\phi)(1-2\phi-4\phi^{2})^2\;,
\ee
and for $n=3$
\be
{V}_{\sin}^{3,1/2}(\phi)=\frac12(1-\phi^2)\;(1-\phi)^2
\ee
\be
{V}_{\sin}^{3,3/2}(\phi)=\frac{1}{18}(1-\phi^2)\;(1-\phi)^2\;(1+2\phi)^4
\ee
\be
{V}_{\sin}^{3,5/2}(\phi)=\frac{1}{50}\;(1-\phi^2)(1-\phi)^2(1-2\phi-4\phi^{2})^4\;,
\ee

In Fig.~4, we plot some of the potentials as a function of the scalar field $\phi$. Also,
the static solutions in this case are given by
\be\label{solsin2}
{\phi}_k(x)=\cos\left(\frac{\theta_n(x)}{a}+ \frac{(k-1)\;\pi}{a}\right)\;,
\ee
where $k$ is integer, which produces different solutions only if $1\leq k\leq 2a$; recall that $\theta_n(x)$ is given by \eqref{theta}.

We notice that these potentials are non negative only for $\phi\geq-1$, and the zero at $\phi=-1$ is not a minimum. Thus, all the static solutions given by Eq.\eqref{solsin2} which interpolate between consecutive minima are topological, and the other, for $k=a+1/2$, which involves the zero at $\phi=-1$ and the neighbor minimum, is non-topological.

\begin{figure}[ht]
\vspace{1cm}
\includegraphics[{height=03cm,width=8.5cm,angle=00}]{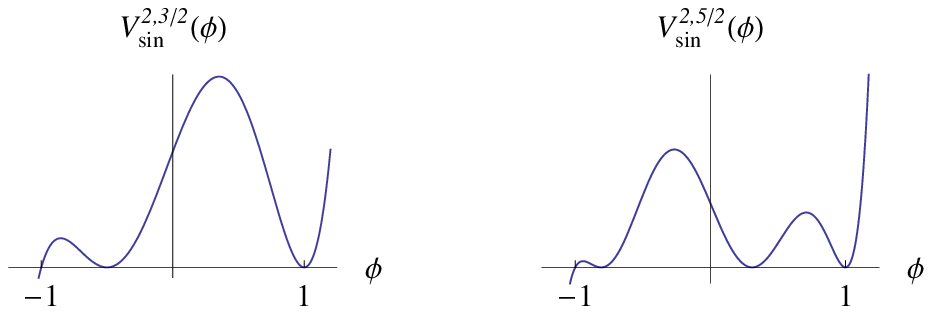}
\includegraphics[{height=03cm,width=8.5cm,angle=00}]{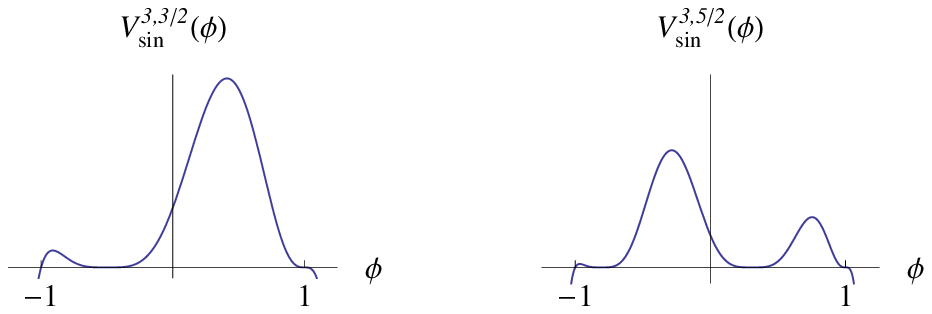}
\includegraphics[{height=03cm,width=8.5cm,angle=00}]{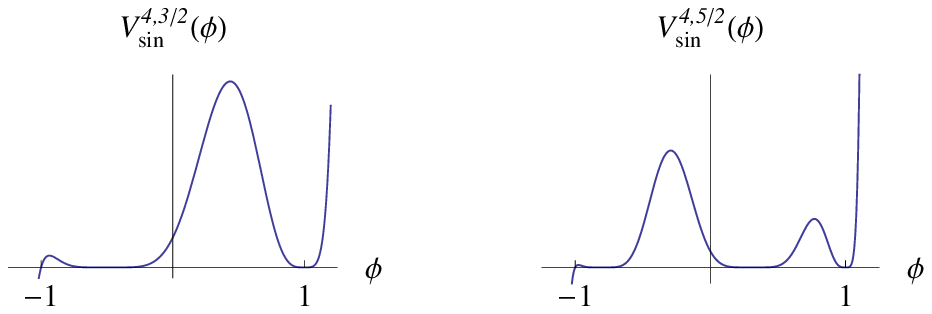}
\caption{Plots of $V^{n,a}_{\sin}$ for some $n$ and $a$.}
\end{figure}

{\bf Superpotential and energy. --} In general, when the potential is non negative, it is possible to introduce superpotential $W=W(\phi)$ such that
\be
V(\phi)=\frac12\left(\frac{dW}{d\phi}\right)^2\;.
\ee
This is the case for the $sine$ family of potentials with $a$ integer. However, In the others cases, for the $cosine$ family of potential with $a$ integer, and for the $sine$ and $cosine$ families of potential with $a$ half integer, the potentials are not non negative. Nevertheless, we can follow the lines of \cite{ablm} and introduce superpotential, for both topological and for non topological solutions. In the case of the $sine$ families, for $a$ integer or semi-integer, the superpotentials
can be written in terms of Chebyshev polynomials if we fix the integer $n$, for instance:
\bes
\be
W^{2,a}_{sin}(\phi)=\frac{(a^2(1-\phi^2)-2)T_a(\phi)
-2a\phi(1-\phi^2)U_{a-1}(\phi)}{a^2(a^2-4)}\,
\ee
for $a\neq2$, and
\ben
W^{3,a}_{sin}(\phi)&=&\frac{\phi\sqrt{1-\phi^2}+\arcsin\phi}{4a}+\frac{\sqrt{1-\phi^2}}{8a^2(a^2-1)}\nonumber\\
&&\times\bigl[2a\phi T_{2a}(\phi)-(1-2a^2(1-\phi^2))U_{2a-1}(\phi)\bigr]\nonumber\\
\;\;\;
\een
\ees
Also, if we fix $a$ we can get other expressions for $n$ integer, for instance:
\bes
\be
W_{\sin}^{n,2}(\phi)\, =\, \frac{2^{n-2}\phi^n}{n} \, _2F_1\left( \frac{n}{2},-\frac{n}{2}, \frac{n}{2}+1,\phi^2\right)
\ee
\be
W_{\sin}^{n,3}(\phi)\, =\, \frac{(-1)^{n+1}}{3}\, F_1\left(\frac{1}{2},-\frac{n}{2},1-n,\frac{3}{2},\phi ^2,4 \phi ^2\right)
\ee
\ees
where $F_1$ is the Appell function of first type.
Also, in the case of the $cosine$ families, for $a$ integer or semi-integer,
the expressions for the superpotentials are similar to the corresponding expressions for $W_{\sin}$, for instance:
\bes
\ben
W^{2,a}_{cos}(\phi)&=&\frac{1}{a^2(a^2-4)}[(a^2(1-\phi^2)-2)\sin(a\;{\arccos}\phi)\nonumber\\
&&+2a\phi\sqrt{1-\phi^2}\cos(a\;{\arccos}\phi)]\;,
\een
for $a\neq2$, and
\be
W_{\cos}^{n,2}(\phi)\, =\, \frac{(-1)^{n+1}}{2}\,   \phi \,
   F_1\left(\frac{1}{2};-\frac{1}{2},1-n;\frac{3}{2
   };\phi ^2,2 \phi ^2\right)
\ee
\ben
W_{\cos}^{n,3}(\phi)&=&\frac{(-1)^{n+1} 3^{n-3} \phi ^n}{n
   (n+2)}\bigg(4 n \phi ^2\nonumber\\ && \times
   F_1\left(\frac{n}{2}+1;-\frac{1}{2},1-n;\frac{n}
   {2}+2;\phi ^2,\frac{4 \phi ^2}{3}\right)\big.\nonumber \\ && \left. +3 (n+2)
   F_1\left(\frac{n}{2};-\frac{1}{2},-n;\frac{n}{2}
   +1;\phi ^2,\frac{4 \phi ^2}{3}\right)\right)\nonumber \\
   \;
\een
\ees

In general, the superpotential simplify the calculations. Particularly, for the energy associated with the corresponding static solutions we can write, for the kinklike solutions,
\be
E^{n,a,k}(\phi_k)=|W^{n,a}(Z_{k})-W^{n,a}(Z_{k+1})|\;,
\ee
and for the lumplike solutions \cite{ablm}
\be
E^{n,a,k}(\phi_k)=2\;|W^{n,a}(Z_{k})-W^{n,a}(\phi(x=0))|\;.
\ee

The general superpotential, for arbitrary $n$ and $a$, is very hard to be calculated. However, in order to calculate the energy, we can circumvent this problem following the alternative route: the kinklike solution $\chi_n(x)$, Eq.\eqref{Fn}, can be approximated by
\be\label{achin}
\chi_n(x)=\pm\tanh(\alpha_n x)\;,
\ee
where the width of the solution is controlled by the inverse of
\be\label{alpha}
\alpha_n=\frac{3\pi}{8}\frac1{n^{0.28}}.
\ee
This is an approximation, but it leads to an error less than two percent for the energy of the corresponding solution.
Also, we can write the energy as
\be
E_n=\int^{\infty}_{-\infty}\left(\frac{d\chi_n}{dx}\right)^2dx\;
\ee
With this last expression, we use \eqref{achin} to get that $E_n=4\alpha_n/3$. Now, comparing the approximated energy value with those from \eqref{eny},
we obtain a good fit by means of \eqref{alpha}. Particularly, for $n=3/2, 5/2$ and $3$ we obtain $\alpha_{3/2}=1.052,$ $\alpha_{5/2}=0.912,$ and $\alpha_3=0.866$.
For the models described by the potential \eqref{upot}, the energy of the defect solution decreases with increasing $n$, but the corresponding width increases.
Finally, for $n\neq 2$, the energies of the static solutions given by Eqs.~\eqref{solsin1}, \eqref{solcos1}, and \eqref{solsin2}, are obtained straightforwardly
from the energy for $n=2$, with the answer $E^{n,a,k}=\alpha_n\;E^{2,a,k}$.

{\bf Final comments. --} In this work, we have introduced and solved two distinct new families of models. The first family is defined in \eqref{upot}, and it is controlled by the single parameter $n$, integer or half integer. The second family is introduced by the deformation function \eqref{def}, which depends of two new parameters, $a$ and $m$. With this deformation we could find new models, depending on $m$ being integer or half integer, and on $a$ and $n$. In this case, we have obtained a diversity of models and their respective defect solutions. We have also found a way of presenting the energy of the defect solutions, both exactly or approximately, when the exact result is hard to be obtained. As one knows, we can also investigate linear stability of the scalar field solutions. Although this issue is out of the scope of the present work, it can be done following the lines of \cite{bglm}.

We would like to thank CAPES, CNPq (Brazil) and DGICYT (Spain) for partial financial support.

\end{document}